\documentclass{emulateapj}

\shorttitle{The Submillimeter Array}
\shortauthors{Ho et al.}

\begin{document}


\title{The Submillimeter Array}


\author{Paul T.P. Ho}
\affil{Harvard-Smithsonian Center for Astrophysics, 60 Garden Street,
    Cambridge, MA 02138\\
    Academia Sinica Institute of Astronomy and Astrophysics,
P.O. Box 23-141, Taipei 106}

\author{James M. Moran}
\affil{Harvard-Smithsonian Center for Astrophysics, 60 Garden Street,
    Cambridge, MA 02138}

\and

\author{Kwok Yung Lo}
\affil{National Radio Astronomy Observatory, 520 Edgemont Road,
Charlottesville, VA 22903\\
Academia Sinica Institute of Astronomy and Astrophysics,
P.O. Box 23-141, Taipei 106 }
\email{ho@cfa.harvard.edu, jmmoran@cfa.harvard.edu, flo@nrao.edu}


\begin{abstract}

The Submillimeter Array (SMA), a collaborative project of the Smithsonian
Astrophysical Observatory (SAO) and the Academia Sinica Institute of Astronomy
 and Astrophysics (ASIAA), has begun operation on Mauna Kea in Hawaii.  A total
 of eight 6-m telescopes comprise the array, which will cover the frequency 
range of 180-900 GHz.  All eight telescopes have been deployed and are 
operational. First scientific results utilizing the three receiver bands at
230, 345, and 690 GHz have been obtained and are presented in the accompanying 
papers. 

\end{abstract}




\section{Introduction}

The Submillimeter Array (SMA) Project was conceived at the Smithsonian 
Astrophysical Observatory in 1983 as a part of a broad initiative by its new
director, Irwin Shapiro, to achieve high resolution 
observational capability across a wide range of the electromagnetic spectrum.
  The aim of the SMA is to use interferometric techniques to explore 
submillimeter wavelengths with high angular resolution.  One measure of the
importance of the submillimeter
window derives from the fact that the bulk of the universe is at a relatively 
cold temperature of about 10K, thereby placing the peak of the radiation curve in 
the submillimeter and far-infrared range.  From the ground, the submillimeter 
wavelengths are as close as we can get to this radiation peak, and high 
resolution observations are only possible from the ground until space
far-infrared interferometry becomes feasible. Furthermore, many more unique 
and high excitation molecular lines become available in the submillimeter 
window.  In the 1980s, only single aperture instruments such as the 10-m 
Caltech Submillimter Observatory (CSO), the 15-m James Clerk Maxwell Telescope 
(JCMT), the 10-m Heinrich Hertz Submillimeter Telescope (SMT), and the 3-m 
Kolner Observatorium fur Submillimeter Astronomie (KOSMA) were operating or 
planning to operate at submillimeter wavelengths.  The SMA was designed to 
increase the available angular resolution by a factor of 30.  A formal 
proposal was presented to the Smithsonian Institution in 1984 \citep{mor84}, 
and was 
reviewed favorably by the community.  Initial funding of the project began with
 the establishment of a submillimeter wavelength receiver laboratory at SAO in
 1987.  Two years later, the design study for the SMA \citep{mas92} was funded, 
 and construction funds followed in 1991.  Several sites for the array were
considered including Mount Graham in Arizona, a location near the South 
Pole, and the Atacama desert in Chile \citep{raf92}.  Mauna Kea in Hawaii was
ultimately chosen, in part due to the existence of good infrastructure 
as well as other programmatic reasons, and 
in 1994 an agreement was 
reached with the University of Hawaii for the construction of the array in  
``Millimeter Valley'' adjacent to the CSO and the JCMT at an 
elevation of 4,080 m.  This specific location was selected because of the 
potential for linking up with CSO 
and JCMT as well as the ability to achieve baselines of at least 500 
m without great changes in elevation, with the possibility of even longer 
baselines in the future. 

The initial concept for the SMA was six 6-m telescopes, the parameters of which 
were driven by: (1) the desire for a fast instrument that would sample the uv 
plane adequately, even for short observations, (2) the desire to have at least 
the same collecting area as the existing submillimeter telescopes, (3) the 
desire for a reasonably large primary beam at the highest frequencies in
order to make the pointing requirements manageable and the field of view 
reasonably large, and (4) the balance between the cost of telescopes, which 
scales faster than the square of the aperture, and the cost of receivers,
 electronics, and correlator.

In 1996, the Academia Sinica Institute of Astronomy and Astrophysics joined the 
 SMA project by agreeing to add two more telescopes and all associated 
electronics, including a doubling of the correlator.  The Academia Sinica agreed
with the cost effectiveness of this plan and funded the expansion of the SMA as
the first astronomical project at ASIAA.  The addition of two 
 elements increased the number of instantaneous baselines from 15 to 28, nearly 
 doubling the speed of the array for some applications.  While the addition of 
 a partner introduced complications in the construction process, the power of
 the SMA and its scientific potential were increased significantly.  

SAO personnel designed the telescope, receivers, electronics, and correlator. 
 Fabrication of the mount, the reflector panels and other subsystems was 
 performed by
 subcontractors, while the assembly was done by SAO staff at the SMA facilities
 located
 at Haystack Observatory in Westford, Massachusetts.  After initial tests 
including holography at 94 GHz and interferometry at 230 GHz, the individual 
telescopes were disassembled, shipped to Hawaii, and reassembled in the SMA 
assembly hall on Mauna Kea.  The Aeronautical Research 
Laboratory (ARL) in Taiwan was the primary contractor for fabrication of the two
 telescopes contributed by ASIAA.  By the end of 2003, all eight elements of 
the SMA had been deployed on top of
 Mauna Kea, and the SMA was formally dedicated on November 22, 2003.  Figure 1
 shows the completed SMA in a fairly compact configuration.  

\section{Description of the Instrument}

The basic characteristics of the SMA are summarized in Table 1.

\subsection{Construction of the Telescopes}

In 1992, SAO chose to be the general contractor for construction of the 
telescopes.  While submillimeter wavelength telescopes had already been built 
 by other institutions, the requirements of interferometry presented additional
challenges such as the need for exceptional mechanical stability and
transportability of the antennas.  The massive size of the mount was necessary 
to provide stability under stringent pointing requirements.  The reflector and 
its backup support structure were designed by Philippe Raffin and the SAO staff 
\citep{raa91, rab91}.  The reflector surface was composed of machined cast 
aluminum panels, which were chosen over carbon fiber panels due to concerns
 over conditions on Mauna Kea that can include windblown abrasive volcanic dust
 and storms that bring heavy snow and thick ice accumulations.  In order to 
achieve the necessary surface accuracy in the manufacturing process, as well 
as good thermal performance in the field, four rows of individual panels, each 
about 1 m in size, were employed.  A mean surface accuracy of six microns 
was achieved for the individual panels.  To hold the surface in shape, carbon 
fiber tubes and steel nodes were used to form an open backup structure. The 
individual panels were attached to the backup structure via four mechanical 
adjusters per panel.  The surfaces were adjusted from the front, as the backup 
structures were covered by an aluminum skin in order to protect them from the 
weather. The adjusters not only held the reflector panels to the backup structure, 
but, because of the over constraint of the four point support, allowed 
twists within each panel to be corrected.  The use of carbon 
fiber tubes provided a lightweight and rigid structure with a very low thermal 
expansion coefficient.  A linear screw drive was chosen to move the telescope 
in the elevation direction because it allowed for a large, stable  walk-in 
cabin for the receivers.  The receiver cabin was built around the mount in 
between the elevation bearings, and the optics chosen were bent-Nasmyth
via a tertiary mirror behind the center of the reflector.  This arrangement 
allowed the 
receivers to be maintained with a constant gravity vector in order to ensure
mechanical stability, which was deemed important for receiver stability.

The telescopes were put together in the SMA assembly hall at the MIT Haystack 
Observatory in Westford, Massachusetts; the first prototype was built in 1996,
 and an improved version was put into operation in 1997.  Two copies of the 
second 
prototype operated successfully as an interferometer at 230 GHz in the fall of
 1998 at Haystack.  These two telescopes were deployed to Hawaii the following 
 year, and obtained first fringes on Mauna Kea in September 1999: a major 
milestone for the project.  From that point on, the production 
models of the telescopes began to arrive in Hawaii, incorporating improvements
in the receiver cabin, electronics, and servo systems.  The final versions of
the primary reflectors of the telescope, including the carbon fiber backup
structures, were assembled on Mauna Kea using a special rotating template.  In 
the last four years,
 the two telescopes from Taiwan arrived in Hawaii, four more telescopes arrived 
 from Massachusetts, and the original two prototype telescopes were rebuilt in
 Westford.  

\subsection {Holographic Adjustments of the Surface}

The reflectors were assembled with an initial accuracy of
about 60 microns. After the reflectors were installed on the telescope mount, 
holographic alignment was performed with a 232 GHz radiation source mounted on 
the catwalk of the Subaru Telescope \citep{sri02}.  Located about 200 m from 
the center of the array at an elevation angle of about 20$^{\circ}$, the test 
signal was observed through the SMA optics with the subreflectors set to their
 near field focus positions.  The far field beam response of the telescope 
under study was
 measured by scanning it while a second telescope provided the phase reference.
Amplitude and phase data were acquired with an on-the-fly technique (i.e., 
continuous 
movement of the reflector).  Fourier transform of the complex map gave the 
distribution of illumination (amplitude) and surface deviations (phase).  The 
scanning resolution was 33$''$, which corresponded to a spatial resolution on 
the reflector of about 10 cm.  The holographic procedure, while simple in 
concept, had many subtleties.  Corrections for the near field geometry, 
multiple reflections, and diffraction effects had to be made.  The effects of 
multiple reflections were mitigated by averaging maps with the subreflectors 
in various focus positions.  Typically, three or four rounds of holography and 
resetting of the panels were required to reach the goal of 12 microns accuracy 
for 
 the surface at an elevation of 20$^{\circ}$.  Figure 2 shows the deviations in 
the surface of one of the reflectors before and after the setting of the surface 
with this holographic 
technique.  The reflector surfaces have been monitored for long-term stability.  
On the 
time scale of several months, the surface accuracy appears to be stable at 
about 11 microns rms.  This includes the changes due to redeployment of the 
telescope from one pad to another.  In the future, holography at 682 GHz will 
be used for more precision, and a celestial holography capability 
will be developed to enable  
studies of the reflector behavior under different conditions of gravitational 
deformation over a wide range of elevation angle.  

The surface accuracy can also be checked by measuring the aperture efficiency 
of the telescope while observing a planet.  Initial measurements indicate 
70-75\% efficiency at 230 GHz, 50-60\% efficiency at 345 GHz, and about 
40\% at 680 GHz.  Optimization of the surface accuracy of each telescope 
is continuing.    

\subsection{Configuration of the Array}

For an interferometer array with a small number of elements, the configuration
is important for achieving the best uv plane coverage and the best image 
quality.  The design of the SMA configuration was driven by the desire for a 
uniform sampling of uv plane spacings within a circular boundary, whose radius sets
the angular resolution.  Scaled Y-configurations such as the VLA are 
centrally condensed and undersample the long spacings.  The redundancies in 
such configurations have distinct advantages, but compromise the uv plane 
coverage if the number of elements in the array is small.  A
 configuration based on the Reuleaux triangle, which is an equilateral triangle 
 whose sides have been replaced with circular arcs with the opposite vertices 
as their centers, was found to provide the most uniform uv plane sampling.  By 
locating the 
interferometer elements on a curve of constant width such as the Reuleaux 
triangle, the maximum separation between telescopes is a constant.  This 
 ensures that the maximum uv plane spacings lie on a circle, thereby resulting in a 
circular beam for observations at the zenith.  The choice of a triangle, 
versus a polygon of more sides, 
ensures that the angles between baseline vectors, and therefore potential 
differences between projected baselines, are maximized.  This results in a more 
uniform sampling distribution in the uv plane.  The sampling of the 
shorter spacings depends on the actual locations of the array elements on the 
curves, which was 
optimized with a neural network search algorithm \citep{ket97}.  Addition of 
the two ASIAA telescopes was accommodated within this Reuleaux triangle scheme 
on an ad hoc basis.  The SMA configuration works well for a small number of 
interferometer elements. For an array with a large number of telescopes, their 
specific
 distribution is less critical.  

To provide different angular 
resolutions the array consists of four nested ``rings" of 24 pads.  Each of 
the ``rings" is an optimized Reuleaux triangle, accommodating up to eight pads.
  The ``rings" are nested in order to share some of the pads and thereby 
reduce costs.  Some compromises were eventually made because of the topography
 of the site. The actual layout of the pads is shown in Figure 3.

Because of the uneven terrain of the SMA, as well as environmental restrictions,
 the telescopes had to be transported 
without the use of rails, as was the case of the VLA.  A special transporter 
was designed to pick up and move the 50-ton telescopes.  
This piece of equipment drives 
under its own power, and is nimble enough so that several antennas can be 
repositioned in a day.

\subsection{Receivers and Electronics}

The front end receiver electronics in each antenna are housed in closed cycle
 helium cryostats (Daikin model CG-308SCPT cryo-cooler).
  The cryostats use a two stage Gifford McMahon system to reach 70 and 15K, 
respectively, and a Joule-Thompson valve to reach 4K.  Each cryostat has room 
for eight receiver inserts and a capacity of 2.5 watts at 4K \citep{blu98}.  An 
optics cage is mounted above the cryostat, which splits the polarizations of 
the incoming radiation via a rotating wire grid and flat mirror assembly.  Each 
 of the two polarizations can be directed into separate receivers.  This 
arrangement allows either a dual-polarization mode for maximum sensitivity 
 or polarization measurements, or the simultaneous operation of a high 
frequency and a low frequency receiver.  Optically injected LO sources are also 
 located above the receivers, while a calibration vane, as well as 
quarter wave plates, can be inserted into the optical path.  The quarter 
wave plates provide circular polarization, but they have only been tested in 
the single receiver mode at 345 GHz.

Three receiver inserts that cover the 230, 345, and 690 GHz bands are 
now in operation.  The heart of these receivers are double sideband mixers 
fabricated with niobium SIS junctions, which are cooled to 4.2K along with the 
second stage HEMT amplifiers.   On the SAO side, these junctions were
fabricated by JPL.  On the ASIAA side, a partnership
with National Tsinghua University, Nobeyama Observatory, and the Purple 
Mountain Observatory produced the junctions \citep{shi02}.  The 
instantaneous bandwidths of the receivers are 
50, 100, and 60 GHz in the 230, 345, and 690 GHz bands, respectively. The double
 sideband receiver temperatures in the laboratory setting were about 2, 2.5, and
 7 times the quantum limit, or about 25, 35, and 200K, respectively 
\citep{blu04}. The double 
sideband receiver temperatures of the operational systems on the Array are 
currently about 80, 100, and 480K, respectively.  Technical descriptions of the
receivers can be found in papers by Blundell et al. (1995), Tong et al. (1996),
and Tong et al. (2002). 
%\citet{blu95},\citet{ton96}, 
%and \citet{ton02}.

The phase-locked LO sources are based on Gunn Oscillators operating in the 100
GHz range, whose signals have been multiplied with diode doublers and triplers. 
  For lower frequencies, the LO is injected into the optical path via a simple 
 mesh grid.  To achieve adequate power at 650 GHz, the LO is injected via a 
Martin-Puplett diplexer.  The LO signals are coupled optically to the mixers from
outside the cryostats.

Because of the large number of mechanical parts that must be tuned within the 
receivers, the goal is to have them under computer control to 
facilitate remote operation of the system and improve operational efficiency.  
The position of the wire grids and mirrors in the optical path, the calibration 
vane, the mechanically tuned LO, the mixers, and the phase-lock loops will all
 be put under servo control.  Much of this capability is already in place for 
the lower frequency bands \citep{hun02}. 

On the telescopes, the receivers have worked quite well, and were able to run
for months and years at a time without failure.  The cryostats must be warmed 
up periodically for coldhead maintenance (every 10,000 hrs.), but the junctions
 themselves are quite robust, despite repeated warming ups and cool downs due
to power failures on site.  

\subsection{Correlator}

The SMA correlator has a flexible hybrid analog-digital design.  After the 
first down conversion, the IF band centered at 5 GHz from each receiver is 
broken up into six contiguous blocks of 328 MHz each, covering a total window 
of 2 GHz.  Each block of 328 MHz, recentered at an IF of 1 GHz, is further 
split into four chunks of 104 MHz (82 MHz spacing). Thus, a total of 24 chunks
 or basebands are derived from each of two receivers for all eight telescopes. 
  A maximum of 384 basebands are therefore presented to the digital part of the 
 correlator, for a maximum of 1344 multi-lag cross correlations.  The digital 
part of the 
 correlator consists of 90 boards, each with 32 custom designed correlator 
chips.  These correlator boards were built by MIT-Haystack as part of the MK 
IV correlator project \citep{whi04}.  Hence, a minimum of two chips can be 
devoted to each baseband correlation.  With 512 lags per correlator chip, a 
data rate demultiplex factor of four (the correlator clock rate is 52 MHz), and
 a factor of two for calculating both 
amplitude and phase, 128 spectral channels are obtained per baseband.  Thus, 
if the full bandwidth is covered, a spectral resolution of 812.5 kHz is 
obtained.  For full polarization measurements with all four stokes parameters,
 the 
spectral resolution would be a factor of two worse or 1.625 MHz.  If fewer 
numbers of basebands are processed, more correlator chips can be used per 
baseband to achieve higher spectral resolutions. For example, if only 
one baseband per block is processed, 16 chips can be used on each baseband, 
achieving 101.6 kHz resolution.  Furthermore, different basebands can be 
processed with different spectral resolutions, and the individual blocks can 
be tuned to different positions within the passband.   By reprogramming the 
correlator boards to put more chips on a baseband, even higher spectral 
resolutions can be achieved.  

When the SMA is linked to JCMT and CSO for joint interferometry beginning in 
2005, the correlator will be able to process the full bandwidth on all 45
baselines with one receiver.  Dual band capability could be achieved by 
reducing the number of baselines or bandwidth.

\section{Calibration of the Instrument}

In order to operate the SMA as an interferometer, many system performance
 calibrations must be done.  As previously described, the surfaces of the 
telescopes are set 
by holographic measurements, and the efficiencies of the telescopes are checked 
by observing planets and their satellites.  The pointing models for the 
telescopes are
determined first with data from optical guidescopes mounted behind holes 
in the primary 
reflectors \citep{pat00, pat04}.  Typically, positions of  more than a 
hundred stars are 
measured throughout the sky, and  19-parameter pointing models are determined in 
order to correct for the collimation, tilt, sag, and encoder offsets.  The 
residuals after the fit are typically 1-2$''$ rms in each axis.  While the 
pointing is stable on the order of days, there are long-term drifts in the tilt 
components of the telescopes, which might be associated with the stability of 
the antenna pads.  After the optical pointing models have been determined, the 
alignment of the radio and optical axes of each telescope is checked by radio 
pointing on planets.  During observations, pointing of the telescopes is 
verified and further improved by measurements of nearby strong continuum sources.  
Pointing measurements can also be made interferometrically by noding the
antennas and analyzing the changes in fringe amplitude.  The coordinates of the 
array elements are determined from the visibility phase
 measurements on strong quasars tracked over wide ranges of hour angle.  
Typically, the baseline data are taken at 230 GHz, and the antenna locations 
can be determined to an accuracy of 0.1-0.2 of a wavelength, or about 0.2
 mm.  Finally, the gain and phase of the array are tracked in real time 
by observing a nearby quasar interleaved with the program sources.  Flux and 
passband calibrations are also done in standard fashion by observing planets. 

Under optimal sky conditions, these calibration procedures may be sufficient.  
However, at higher frequencies and during poor weather conditions, auxiliary 
techniques will have to be implemented to correct for the gain and phase 
fluctuations.  A number of different techniques are being considered, 
including: self calibration when the sources are strong enough and simple 
enough; calibration with respect to a maser source if the frequencies of the 
lines are nearby; cross calibration between different receivers utilizing 
quasars at lower frequencies or masers; and measuring 
phase fluctuations by monitoring the atmospheric water lines at 183 GHz 
\citep{wie01} or 20$\mu$ \citep{nay02}, or the total power from the sky 
\citep{baa04,bab04}.  These techniques, which will be important in the 
future for the operation of the Atacama Large Millimeter Array (ALMA), 
currently under construction in Chile, are also actively being developed by 
other groups (e.g. Welch 1999).

\section{Array Performance}

During the last twenty years, millimeter wavelength interferometry has become 
a well developed field \citep{sar93}, and the SMA will push this research to 
wavelengths 
shortward of 1 mm with angular resolution better than an arcsecond.  While 
the array is just now being completed, early results show that: (1) a surface 
accuracy of the telescopes of about 12 microns can be achieved, (2) the 
absolute pointing accuracy at the level of 2 arcseconds rms can be achieved by 
frequent monitoring of a nearby calibrator, (3) the receivers are sensitive and 
are operating at less than about seven times the quantum limit on the 
telescopes, (4)
 the correlator works properly, and (5) amplitude and phase stability are good 
 and are easily corrected with nearby calibrators on timescales of 30 minutes 
under favorable weather conditions.  We have also learned that good weather 
with low opacity is a precious commodity that must be exploited effectively
through the use of dynamic scheduling. 

Figure 4 shows the first image made with all eight elements of the SMA 
operating at the J= 2 - 1 CO line of Mars.  Comparisons of results from the SMA
with other mm-wave interferometers such as IRAM, BIMA, OVRO, and NMA at 230 GHz,
 show that the 
images are consistent in terms of structures and intensities.  The scientific 
potentials are illustrated by the results reported in the accompanying papers. 
 The abundance of spectral lines in the submillimeter window has been 
demonstrated by the observations of IRAS 18089-1732 \citep{bea04, beb04} and 
IRAS 16293-2422 \citep{kua04}.  The vertical abundance of CO and HCN in the 
atmosphere of Titan has been measured \citep{gur04}.  A circumbinary disk has 
been imaged in L1551 \citep{tak04}.  The nearest circumstellar disk in TWHya 
has been imaged with about 100AU resolution \citep{qic04}.  The first 690 GHz 
interferometer maps were obtained towards IRC10216 \citep{you04}.  The first 
 submillimeter image with subarcsecond resolution of the Orion K-L region was 
obtained 
\citep{bec04}. High velocity as well as low velocity outflows have been 
detected in V Hya \citep{hir04}.  Polarized SiO maser features were imaged in 
VY Canis Majoris \citep{shi04}. The ultracompact HII region G5.89 has been 
demonstrated to be the source of a molecular outflow imaged in the SiO 
line \citep{sol04}.  By surveying a number of other bright ultracompact HII 
regions, some have been found to be suitable calibrator sources \citep{suy04}. 
 Nearby galaxies M51 \citep{mat04} and M83 \citep{sak04} have been imaged with 
 small mosaic maps.   Interacting galaxies have also been studied in the VV114 
system \citep{ion04} and the NGC6090 system \citep{wan04}.  

{\bf Acknowledgments.}  We thank the Smithsonian Institution and the Academia 
Sinica for their support of this project.  We thank the entire
SMA team from SAO and ASIAA for their hard work over many years 
in making this instrument a
 reality.  The most gratifying thanks are coming in the form of the
exciting scientific results that this instrument is beginning to produce.

\clearpage

\begin{table*}[t]
\begin{center}

\centerline{Table 1. Basic Characteristics of the SMA}
\bigskip
\bigskip

\begin{tabular}{ll}
\hline 
\hline
\bigskip
Components                      & Specifications                         \\    
\hline

Interferometer elements         & 8 6-meter, f/0.4 paraboloids, bent-Nasmyth 
optics  \\
Telescope mount                 & alt-azimuth \\
Telescope backup structure      & carbon fiber struts, steel nodes, rear 
cladding \\
Primary reflector               & 4 rows of 72 machined cast aluminum panels\\
Surface accuracy                & 12 microns rms \\
Secondary reflector             & machined aluminum, 10Hz chopping\\
Array configuration             & 4 nested rings, 24 pads, up to 8 pads per 
ring\\
Available baselines             & 9 - 500 meters\\
Operating frequencies           & 180 - 900 GHz\\
Maximum angular resolution      & 0.5$''$ - 0.1$''$\\
Primary beam field of view      & 70$''$ - 14$''$\\
Receiver bands                  & 230, 345, 460, 690, and 850 GHz\\
Number of receivers             & 8 per telescope, 2 simultaneous bands\\
Correlator                      & hybrid analog-digital, 2 x 28 baselines\\
Number of spectral channels     & 172,000\\
Maximum bandwidth               & 2 GHz\\
Maximum spectral resolution     & 0.06 MHz\\
Maximum data rate               & $>$ 10 GB/day for 1-second integrations\\

\hline
\end{tabular}
\end{center}
\end{table*}

\clearpage

\begin{figure}
%\plotone{f1.eps}
\includegraphics{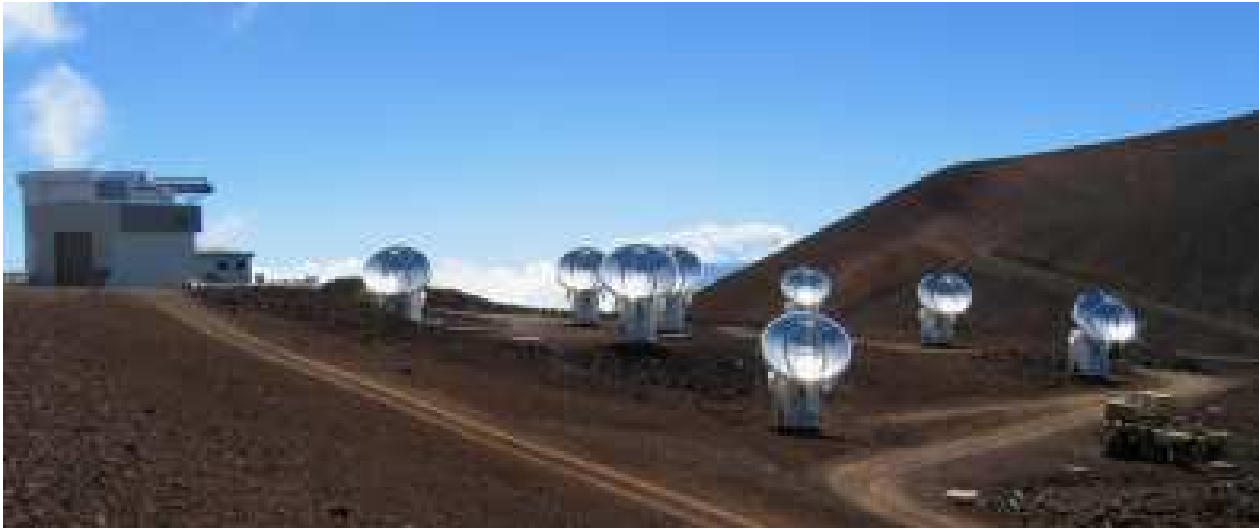}
\vskip4.0in
\caption{View of the SMA in the direction of Mauna Loa.  The assembly/
maintenance building and attached control building are in the top left.  The 
JCMT can be seen rising above them in the background.  The slope of Pu'u 
Poli'ahu rises in back of the Array on the right side.  The transporter used 
to move the antennas is in the right side of the foreground.  In its most 
compact configuration, all antennas can occupy the flat plateau where four of 
the antennas sit.  At the end of 2003, all eight elements of the SMA were 
operating on Mauna Kea in Hawaii. 
Ray Blundell and Bill Liu headed the antenna groups at SAO and ASIAA/ARL, 
respectively.  
\label{fig1}}
\end{figure}

\clearpage

\begin{figure}
%\plotone{f2.eps}
\includegraphics{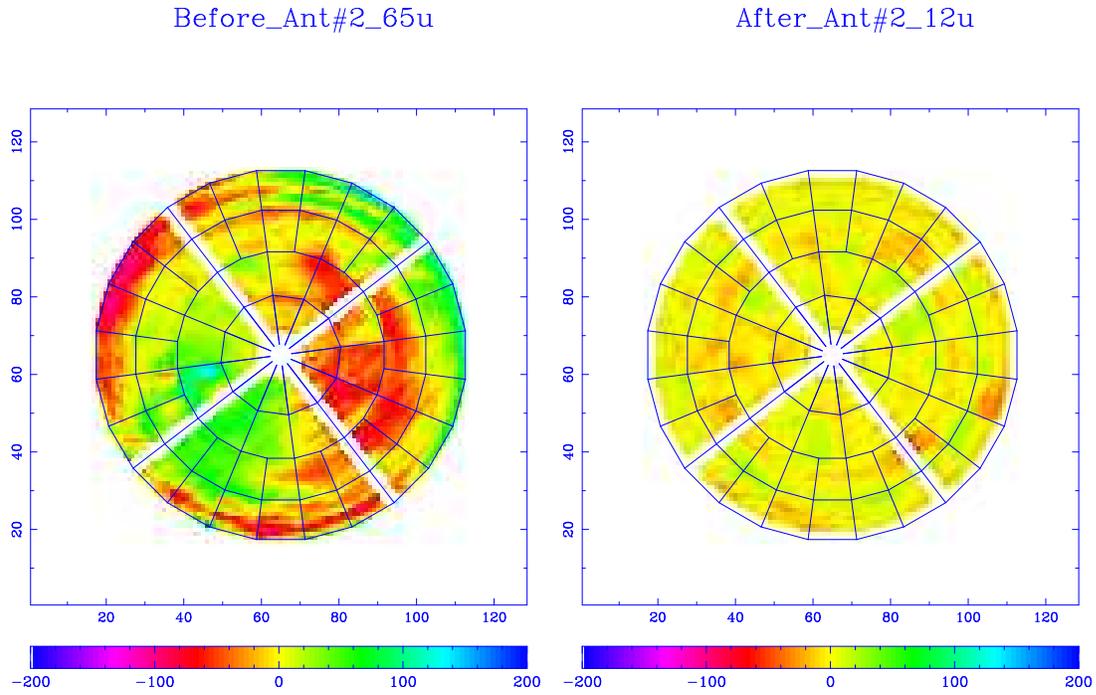}
\vskip4.0in
\caption{Holographic measurements of the surface of Telescope No. 2, before 
(left) and after (right) a series of panel adjustments.  The rms accuracy 
improved from about 65 microns to about 12 microns. T.K. Sridharan and Nimesh 
Patel have been leading the efforts to measure and reset the surfaces of all 
the telescopes.  
\label{fig2}}
\end{figure}

\clearpage

\begin{figure}
%\plotone{f3.eps}
\includegraphics{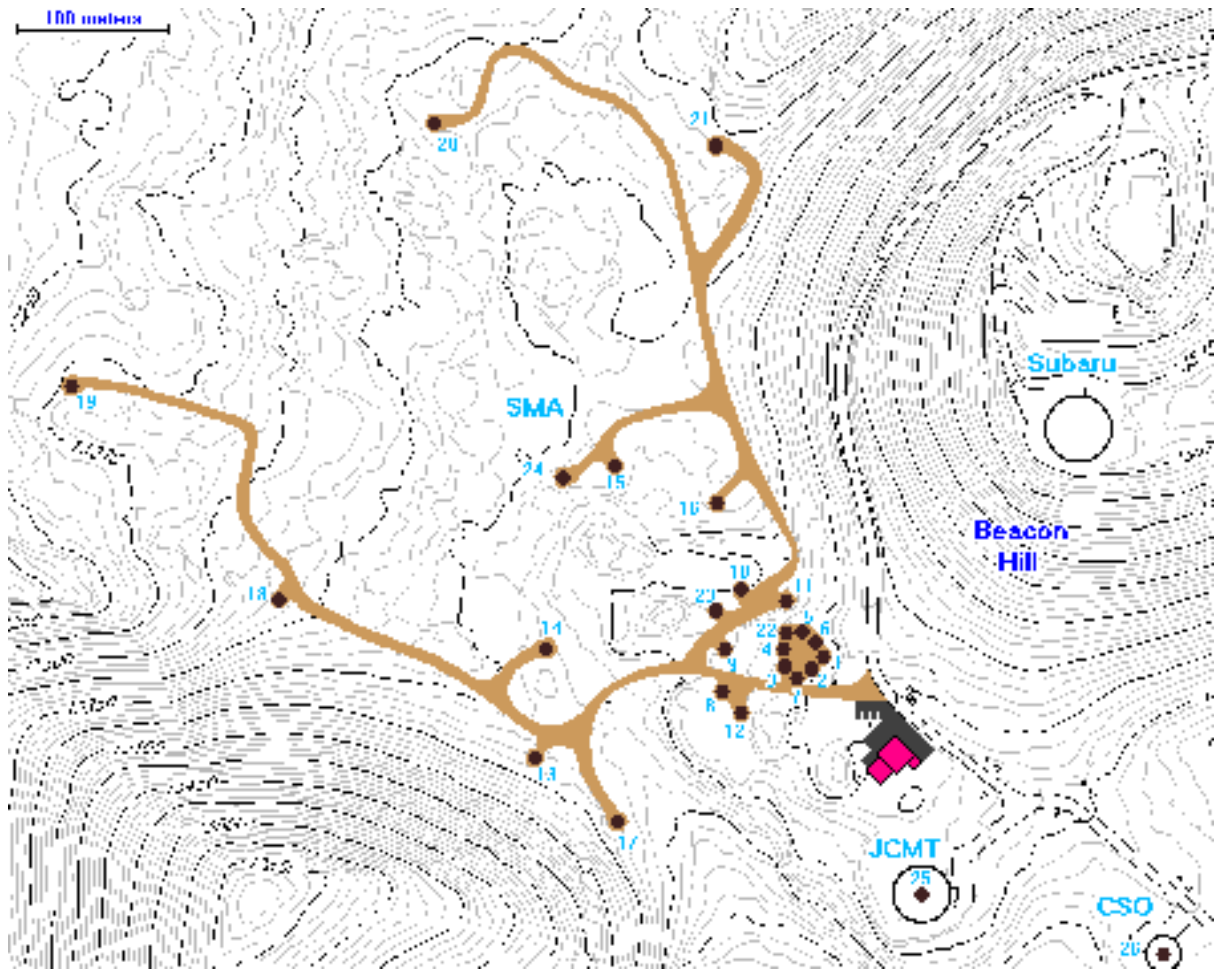}
\vskip6.0in
\caption{The 24 pads of the SMA are distributed in 4 nested rings.  The design 
followed the Reuleaux triangle pattern as much as possible, as constrained by 
the site. Eric Keto was responsible for optimizing the configuration.  Ken 
Young produced this diagram.  
\label{fig3}}
\end{figure}

\clearpage

\begin{figure}
%\plotone{f4.eps}
\includegraphics{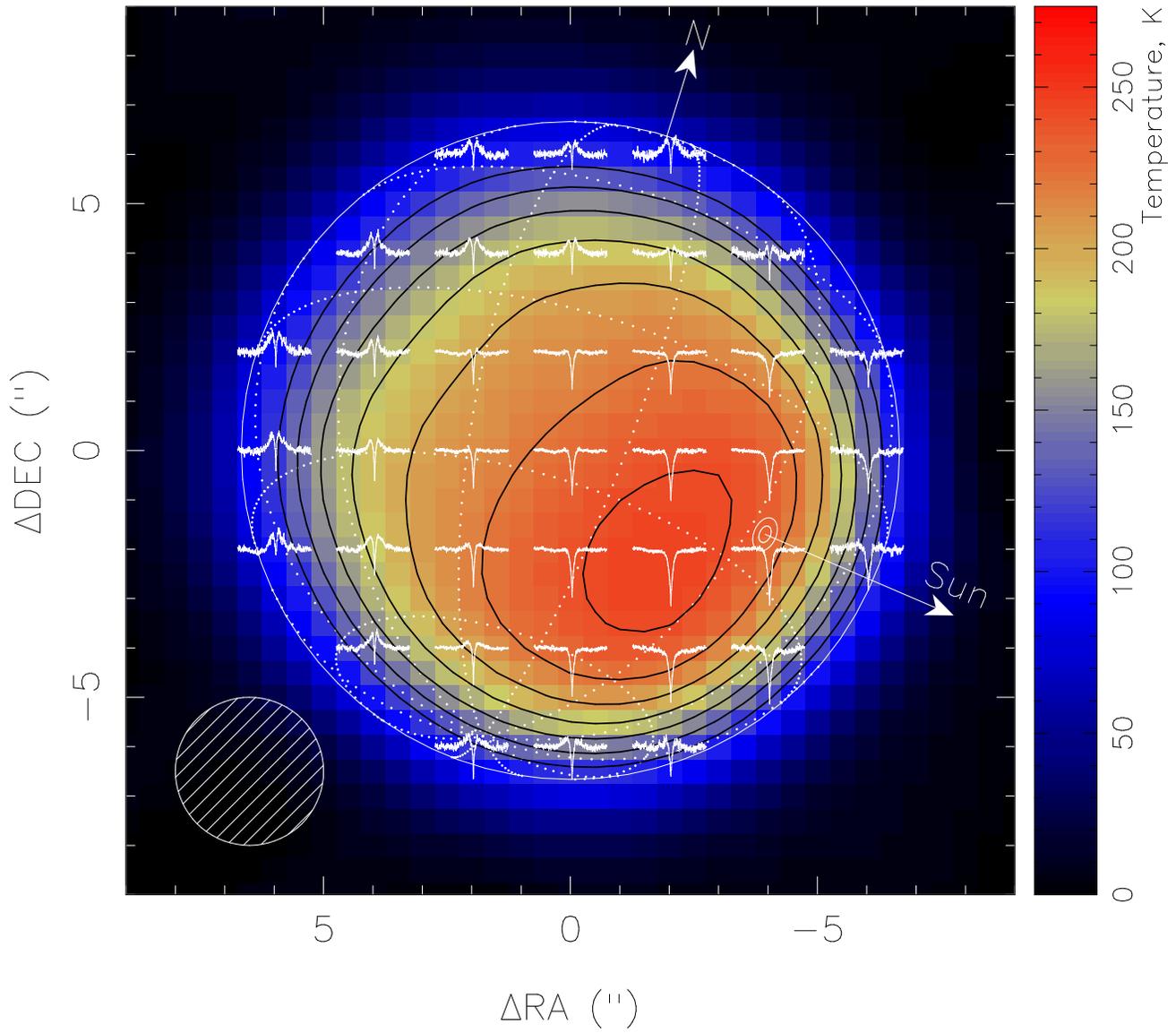}
\vskip6.2in
\caption{The first image obtained with the full eight element array on 12 
November 2003 of the 1.3 mm thermal surface emission and CO(2-1) atmospheric 
absorption from Mars at a spatial resolution of 3".  At the time of 
observations the apparent diameter of Mars was 13.3".  The absorption line 
profiles are pressure broadened and can be used to infer the vertical 
distribution of CO and temperature in the atmosphere.  This image was made by 
Mark Gurwell.
\label{fig4}}
\end{figure}

\clearpage
\end{document}